\documentclass[tighten]{emulateapj}

\slugcomment{Preprint, \today}

\shorttitle{Arctic Infrared Sky}
\shortauthors{Steinbring}

\def\plotone#1{\centering \leavevmode
\epsfxsize=0.90\columnwidth \epsfbox{#1}}
\def\plotonenarrow#1{\centering \leavevmode
\epsfxsize=0.55\columnwidth \epsfbox{#1}}
\def\plottwo#1{\centering \leavevmode
\epsfxsize=1.95\columnwidth \epsfbox{#1}}
\def\plottwomedium#1{\centering \leavevmode
\epsfxsize=1.90\columnwidth \epsfbox{#1}}

\begin{document}

\title{Thermal Infrared Sky Background for a High-Arctic Mountain Observatory}

\author{Eric Steinbring\altaffilmark{1}}

\altaffiltext{1}{National Research Council Canada, Herzberg Astronomy and Astrophysics, Victoria, BC V9E 2E7, Canada}

\begin{abstract}

Nighttime zenith sky spectral brightness in the $3.3-20~\mu{\rm m}$ wavelength region is reported for an observatory site nearby Eureka, on Ellesmere Island in the Canadian High Arctic. Measurements derive from an automated Fourier-transform spectrograph which operated continuously there over three consecutive winters. During that time the median through the most transparent portion of the $Q$ window was $460~{\rm Jy}~{\rm arcsec}^{-2}$, falling below $32~{\rm Jy}~{\rm arcsec}^{-2}$ in $N$ band, and to sub-Jansky levels by $M$ and shortwards; reaching only $36~{\rm  mJy}~{\rm arcsec}^{-2}$ within $L$. Nearly six decades of twice-daily balloonsonde launches from Eureka, together with contemporaneous meteorological data plus a simple model allows characterization of background stability and extrapolation into $K$ band. This suggests the study location has dark skies across the whole thermal infrared spectrum, typically sub-$200~\mu{\rm Jy}~{\rm arcsec}^{-2}$ at $2.4~\mu{\rm m}$. That background is comparable to South Pole, and more than an order of magnitude less than estimates for the best temperate astronomical sites, all at much higher elevation. Considerations relevant to future facilities, including for polar transient surveys, are discussed.

\end{abstract}


\section{Introduction}\label{introduction}

Coordinated, efficient wide-field surveys spanning extended monitoring periods are needed to catch the shortest, faintest, and rarest transient phenomena. One such search is for the electromagnetic counterparts of binary black-hole mergers now being detected via gravitational waves \citep{Abbott2016}. Multiwavelength follow-up of these and neutron-star kilonovae are underway worldwide from the radio through to the optical, as well as with X-rays and $\gamma$-rays from space \citep[e.g.][]{Kasliwal2016}. Viewing those sources in the thermal infrared is desirable as they are less obscured by intervening dust, and if done from a cold, high-latitude observatory this could provide both simultaneous coverage during seasonal dark periods together with low local background - including that from the telescope - which is expected to be a powerful combination \citep{Yuan2013}. 

In winter, polar sites enjoy high clear-sky fractions and typically lie inside encircling upper-atmopheric winds, called polar vortices, that isolate their frigid air - which soon desiccates. Furthermore, under continuous darkness for months, a radiative surface condition induces a strong and stable low-altitude thermal inversion, effectively trapping what little cloud cover persists; thin aerosol attenuation due primarily to suspended ice crystals: ``diamond dust" \citep{Lawrence2004, Steinbring2012}. This stratified atmosphere provides excellent seeing when above surface effects on the central Antarctic plateau \citep[][and references therein]{Saunders2009} and from elevated terrain in the Arctic \citep{Hickson2013, Steinbring2013}. So the only significant remaining hindrance to infrared polar survey cadence and depth is thermal emission of the atmosphere itself.

Near the Eureka weather station on Ellesmere Island, Canada, at $80^{\circ}$ North latitude, sea-level air temperatures in mid-winter can get to $-50~{\rm C}$ and are typically $-40~{\rm C}$ to $-30~{\rm C}$ \citep[reviewed in][]{Steinbring2010}. During the continuous night from October through February a steep (positive) lapse rate of $10$ to $20~{\rm C}~{\rm km}^{-1}$ is usually maintained. Thus, for the Polar Environment Atmospheric Research Laboratory (PEARL), adjacent to Eureka and at 600 m elevation above sea level (a.s.l.), the air temperature is almost always at its peak within the troposphere, between $-30~{\rm C}$ to $-10~{\rm C}$. Ellesmere Island has greater summits, topping 2600 m, although none match the Greenland icecap height of 2800 m. Analogously for South Pole ($90^{\circ}~{\rm S}$), also at 2800 m elevation, surface temperatures hover around $-60~{\rm C}$ in July, rising to $-40~{\rm C}$ at 300 m above, close to the inversion peak. The central Antarctic glacial plateau reaches even higher at the other ``Domes," offering the most extreme nighttime surface air temperatures on Earth, at times surpassing $-90~{\rm C}$ for Dome A ($80^{\circ}~{\rm S}$) and 4200 m elevation. Despite similar elevation, however, the highest temperate mountain sites are comparatively warm, for example, Maunakea ($20^{\circ}~{\rm N}$, 4200 m a.s.l.) is typically near $0~{\rm C}$ during winter nighttime, with diurnal variation often taking summit air temperatures above freezing during the day.

A thermal sky-emission model is developed here within a suitable range of atmospheric transmission, allowing a High Arctic mountain to be put into context with other polar and mid-latitude sites.  Data from balloon-borne radiosondes launched at Eureka then constrain sky-brightness temperature statistics. Those are verified with spectroscopic observations from the ground obtained with an Atmospheric Emitted Radiance Interferometer (AERI), a device essentially identical to one previously operated at Dome C, Antarctica \citep{Walden2005}. The model is described in Section~\ref{model}. Afterwords, in Section~\ref{observations}, archival spectra and contemporaneous sky-clarity estimates are presented, from which emissivities and zenith sky brightnesses are extracted, and global comparisons made. Section~\ref{summary} summarizes results with some discussion of future prospects for all-sky polar infrared synoptic surveys from both hemispheres.

\section{Model and Context}\label{model}

Within its transparent windows, thermal emission from the atmosphere approaches that of a blackbody. Below clear skies, radiant surface emission goes down proportionally with transmission - those photons not radiated to space. This has been successfully applied in characterizing Antarctic plateau sites by assuming saturated air \citep[e.g.][]{Hidas2000}, with the caveat that the inferred effective sky temperature is actually that of the peak in the thermal inversion. What is investigated in this study are potential variation in atmospheric transmission and peak inversion temperature, which imposes limits on Arctic infrared sky brightness. 

Consider downwelling radiation from a uniform horizontal slab of air at temperature $T$ and fixed emissivity $\epsilon$.  Below it, the wavelength-dependent received flux $F$, commonly referred to in astronomy as a spectral ``brightness" is given by the Planck law, in units of ${\rm Jy}~{\rm arcsec}^{-2}$: $$F(\lambda, T)=4.70\times10^{15} {{\epsilon h c}\over{\lambda^3}} \Big{[}\exp{\Big{(}{{h c}\over{k_{\rm B} \lambda T}}\Big{)}}-1\Big{]}^{-1}, \eqno(1)$$ where $h$ is the Planck constant, $c$ is the speed of light in vacuum, and $k_{\rm B}$ is Boltzmann's constant.

\subsection{Transparency Range and Applicability}

Only the thermal component of emission in $K$-band and redder is of interest here. Transparency of the atmosphere in this regime is due primarily to water vapour content, and Eureka is extremely dry, with a mean precipitable water vapour (PWV) in winter of 2 mm \citep{Lesins2009}, and best conditions approaching 1 mm at PEARL \citep{Steinbring2010}. Zodiacal continuum emission makes a relatively small addition to the background near $2.3~\mu{\rm m}$, and a clean cutoff here avoids a plethora of narrow hydroxyl lines blueward. Moonlight is not important except at wavelengths shorter than $2~\mu{\rm m}$. For PEARL, moonless $J$-band ($\sim 1.2~\mu{\rm m}$) sky brightness is found to be similar to midlatitude sites \citep{Sivanandam2013}.

Atmospheric emissivity longward of $2.3~\mu{\rm m}$ is routinely simulated, with various codes available that can generate synthetic spectra. These sum molecular absorption line by line, correcting for pressure and temperature within discrete atmospheric layers and outputting the resulting integrated transmission observable at the ground. That allows for experimentation with vertical atmospheric profiles and the molecular species modelled. Beyond water vapour and ozone, those might include methane, nitrous oxide, carbon monoxide, carbon dioxide, or other trace gases. In this study, a single parametrization with PWV was found sufficient to capture variation in conditions within the passbands considered.

\begin{figure}
\plotone{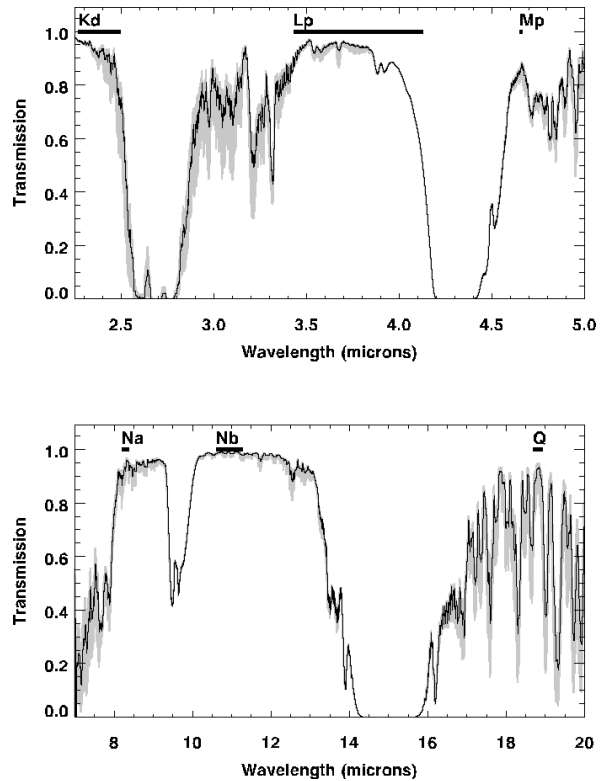}
\vspace{1 mm}
\caption{Bandpasses considered in this study in two broad wavelength regimes. The label ``p" used here has the same meaning as the usual prime symbol. The transmissive properties for a model atmosphere under a range of dryness (grey shading) typical for Gemini North and South, and suitable for comparison to Eureka are shown: PWV between 1.0 mm to 5.0 mm; black curve is for 1.6 mm, the median for Maunakea.}
\label{figure_transmission}
\end{figure}

The ATRAN \citep[Atmospheric TRANsmission;][]{Lord1992} library was employed, obtained from the Gemini website\footnote{http://www.gemini.edu/sciops/telescopes-and-sites}, as used in their online integration-time calculators for infrared instruments. Figure~\ref{figure_transmission} plots a range of transmission at zenith: the black curve is for 1.6 mm of PWV, the median for Gemini North (Maunakea); grey shading shows the range from PWV of 1.0 mm to 5.0 mm. The lower limit serves as a natural choice for a benchmark; the upper limit is near the median at Gemini South (Cerro Pachon; $30^{\circ}$S, 2700 m a.s.l., ${\rm PWV}=4.3~{\rm mm}$), what is considered wet for infrared observations at Gemini North, and extreme for PEARL. Bands corresponding to $K_{\rm d}$ (the optimized ``dark" $K$ filter), $L_{\rm p}$, $M_{\rm p}$, $N_{\rm a}$, $N_{\rm b}$, and $Q$ are indicated. Within these bands, in the absence of aerosols, emissivity for dry air is ideally $\epsilon=0.005$ or 0.5\%, on average approaches 5\%, and apart from the red $L_{\rm p}$-band edge is less than 20\%. These are adopted as a useful range in the comparative analysis that follows.

\subsection{Dependence on Nighttime Temperature}

\begin{figure}
\plotone{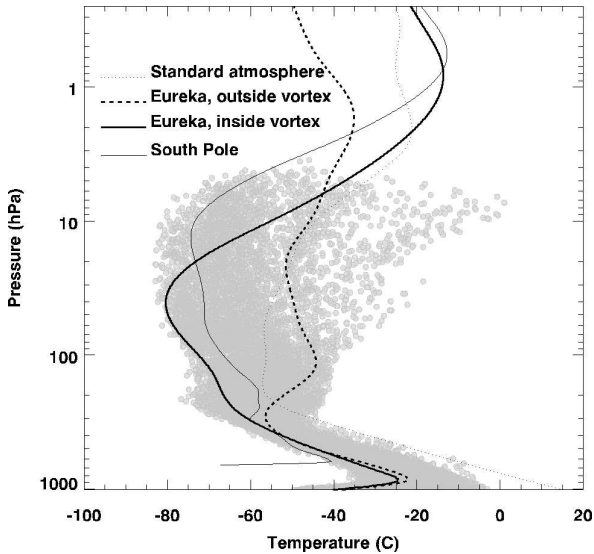}
\vspace{1 mm}
\caption{Temperature measurements above Eureka from balloonsondes (grey cloud of points) during darkness over three full sundown to sunup periods; long-term averages including LIDAR data are shown as thick curves, and in mid-winter from South Pole; dotted curve: U.S. Standard Atmosphere (1976). This last would be more appropriate at mid-latitudes.  All curves have been smoothed by fitting with splines.}
\label{figure_profiles_pressure}
\end{figure}

Regular twice-daily balloon-borne radiosonde launches have been made from Eureka since 1957; Figure~\ref{figure_profiles_pressure} shows a subset of those temperature data, plotted by sampled air pressure - typically at 20 unique levels per flight. The grey cloud of filled circles are all soundings obtained from Eureka for just three winters (2006 to spring 2009) from October 21 through February 20, that period corresponding to the Sun continuously below the horizon. The upper limit to the samples is associated with the highest point in the flight, at which the balloon bursts. The reason for choosing these particular years to display will be evident later, as they coincide with the AERI measurements to follow. Note that during this time, on only one occasion did a measured air temperature rise above $0~{\rm C}$.

One can see from the exponential dependence on $T$ in equation 1 that for an atmosphere stratified into separate (and equally dense) layers of air it would be the layer with the warmest temperature which should dominate. However, those temperatures above the troposhere can be safely ignored for the infrared background discussed here. The hydroxyl emission associated with the upper stratosphere and mesophere at $\sim 1~{\rm  hPa}$ has been explicitly excluded. Also, the relative pressures there are reduced by up to 3 orders of magnitude compared to the ground. For example, if temperatures were uniformly $-60~{\rm C}$ for pressures less than 500 hPa this would constitute just about half of the atmosphere but contribute under 1\% of the downwelling flux at $2.3~\mu{\rm m}$ in equation 1, if the remainder of the atmosphere were at $0~{\rm C}$. This fraction does grow as surface temperature drops and towards longer emitting wavelengths, but the lower layer would still comprise over 63\% of the emission at $20~\mu{\rm m}$, if instead the surface temperature were $-40~{\rm C}$.

Real vertical temperature profiles above Eureka in darkness are more complex, but can be effectively described as intermediate between that of a high-elevation mid-latitude site and winter conditions for South Pole. Seasonal behaviour for South Pole has been well characterized via LIght Detection and Ranging (LIDAR) and high-altitude balloons by \cite{Pan2003}; a thin black curve in Figure~\ref{figure_profiles_pressure} indicates the average profile in July, that is, in austral mid-winter. There is an acute inflection near the surface associated with the thermal inversion. Accordingly, a thick black curve shows the typical condition at Eureka in October through February, for times when it is interior to the northern polar vortex; using spline fits to the profiles reported by \cite{Duck2000} in their Figure 5 (outside the vortex: dashed curve). Duck et al. determined when Eureka fell inside or outside the vortex based on the 10-hPa pressure-level winds. Near-calm conditions characterize those times when Eureka is closest to being centred within the vortex, with strongest winds corresponding to when it lies directly underneath. This has significant impact on the upper atmosphere, but less so near the ground.  Even those times when Eureka is outside the vortex, the peak temperature of the surface-based inversion is raised, on average, only a few degrees. This inversion condition does not always hold though. When it is weak or nonexistent, surface temperatures may reach an extreme similar to that typical of a temperate site; a standard atmospheric profile with a lapse rate of $\gamma = -6.5~{\rm C}~{\rm km}^{-1}$ is shown for comparison. This has a scale height of $7.6~{\rm km}$ yielding a constant ratio $(T_0 - T)/[\gamma \log{(p_0/p)}]$, where $T_0 = 15~{\rm C}$ is the standard sea-level surface temperature, at pressure $p_0$ of 1014 hPa. As this exponential pressure profile seems generally applicable for heights above about $H=3~{\rm km}$ to the tropopause - notice the parallel slopes with log-pressure of the grey points as well as the curve for South Pole - a form with temperature falling as $$T=T_0 - \gamma H \eqno(2) $$ fits data in that regime, where $T_0$ refers to the local temperature at the surface assuming no inversion.

\begin{figure*}
\plottwo{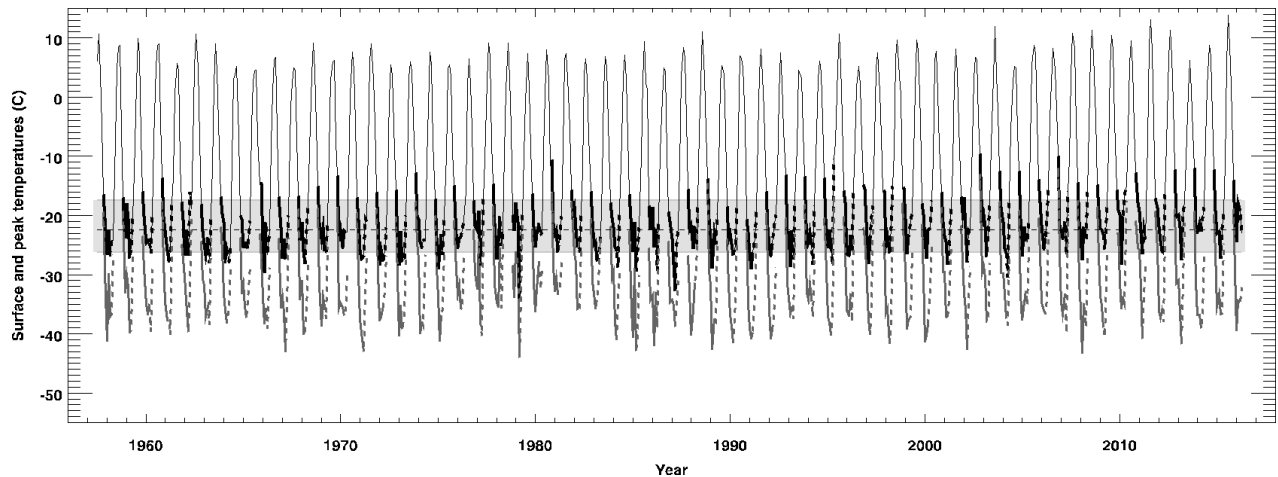}
\vspace{1 mm}
\caption{Monthly mean peak inversion temperatures above Eureka since 1957. The thicker curves indicate months during darkness (dashed curves for two months following sunrise); horizontal dashed lines are the median over those periods and range within 1 standard deviation. Grey curves are temperatures at sea level.}
\label{figure_peak_temperatures_legacy}
\end{figure*}

The usual winter condition for Eureka includes a strong thermal inversion, so it is instructive to look at the inter-season variation and for any longer-term trend in the inversion-peak temperatures that can occur.  Monthly averages in the Figure~\ref{figure_peak_temperatures_legacy} shows the peak temperature in the inversion over the entire set of aerosonde data available for Eureka. The average monthly temperature at the peak of the inversion is shown as a thin black curve.  Those periods during dark months - October through February - are highlighted with a thicker-black overlay; March and April are included as dashed curves.  The same is shown for surface temperatures (dark grey). This helps illustrate the following behaviour: temperatures at the surface and at the peak of the inversion will, on average, drop throughout winter and not increase month-to-month again until well after sunrise.  And despite variation in surface temperatures over past decades, the range of peak temperatures has been remarkably stable. The grey horizontal band is the mean peak temperature, plus and minus one standard deviation: $\bar{T}_{\rm peak}=-22\pm4~{\rm C}$. Other published atmospheric studies over this time period do not report specifically on the inversion peak, but are still in general agreement: for example, no measurable trend in 500 hPa temperatures ($0.11\pm 0.12~{\rm C}~{\rm decade}^{-1}$, 1961 to 2007) is reported by \cite{Lesins2009}.

\subsection{Flux Correction for Thermal Inversion}

\begin{figure}
\plotone{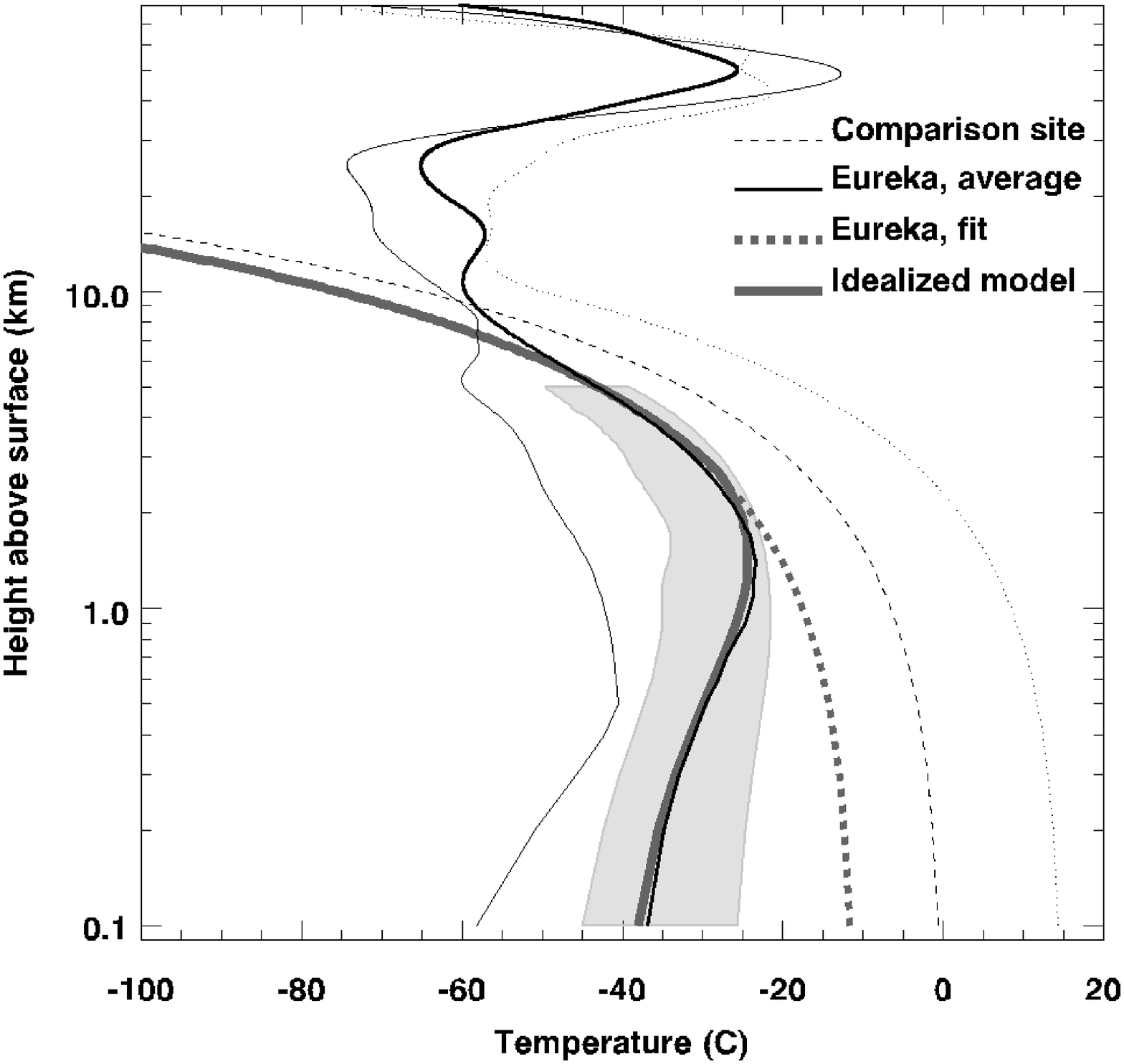}
\vspace{1 mm}
\caption{Same as Figure~\ref{figure_profiles_pressure} except in logarithmic height, normalized by elevation. A scaled standard atmosphere fits the profiles from about 2 km to 10 km above. For the range of peak inversion temperatures and heights that occur at Eureka, equation 3 is a good model; shown as a thick grey curve.}
\label{figure_profiles_height}
\end{figure}

The intent here is allow direct comparison of effective sky temperatures at Eureka with other sites, including mid-latitude mountains. Reforming the plots in Figure~\ref{figure_profiles_pressure} via an exponential pressure profile, and normalizing to the surface elevation of the sites results in the curves shown in Figure~\ref{figure_profiles_height}. Note that, as anticipated, above 3 km from the surface (and below the tropopause) the standard atmospheric profile is suitable. This is shown as a thick-dashed line for Eureka; a thin-dashed curve is for a surface temperature of $0~{\rm C}$, appropriate for Maunakea. Similarly, the summit of Cerro Pachon would have a starting point of $9~{\rm C}$ (not shown). The light grey shading shows the range of temperatures obtained from all Eureka aerosonde profiles as described above; minimum through to mean values in 100 m altitude increments.  Eureka surface temperatures can actually be colder than the peak of the inversion at South Pole, which is usually $20~{\rm C}$ warmer in the first 300 m. The inversion at Eureka can also be of similar strength, but does not typically turn over until about 1 km up in the atmosphere.

The peak temperature is restricted, provided the standard temperature profile holds at higher altitude. Demanding $H_{\rm peak}=(T_0 - T_0^{'})/(\gamma - \gamma^{'})$, where $\gamma^{'}$ refers to the lapse rate within the inversion, and $T_0^{'}$ is the surface air temperature, one can write $$T = T_0^{'} - C^{'} \gamma^{'} H / \exp{(H/H_{\rm peak})}, \eqno(3) $$ and if that meets smoothly with the standard atmosphere at twice $H_{\rm peak}$ (about $2~{\rm km}$ to $3~{\rm km}$ above the surface) the correction $C^{'}$ is given by $e^2 \cdot (1 + \gamma/\gamma^{'})/2 \gamma^{'}$. This is plotted as a thick grey curve in Figure~\ref{figure_profiles_height} for $\gamma^{'} = 15~{\rm C}~{\rm km}^{-1}$ and $T_0^{'}=-40~{\rm C}$, coinciding with the mean profile. Taking things one step further, for the case of $\gamma^{'}\approx (T_{\rm peak}-T_0^{'})/H_{\rm peak}$, equation 3 can be reduced to $T_{\rm peak} \approx T_0^{'} + 3.74~{\rm C}$, when $H_{\rm peak}=1~{\rm km}$. That is consistent with the near-vertical profile towards warmer surface temperature in Figure~\ref{figure_profiles_height}. So even when it is as warm as $-25~{\rm C}$ in Eureka the temperature at the peak of the inversion should still be less than $-21~{\rm C}$.

This simple relationship predicts a narrow distribution of thermal sky background above Eureka during periods when the inversion is present. Note that a difference in temperature up or down by $4~{\rm C}$ would correspond to a scaling by just 20\% in downwelling flux in equation 1 at $5~\mu{\rm m}$. That suggests the following: the strength of the thermal inversion well constrains the downwelling flux, and so for a given range of $\epsilon$ the sky brightness below $H_{\rm peak}$ is readily recovered via $F(\lambda, T_{\rm peak})$, and conversely, measuring the sky brightness and finding the peak temperature from the radiosonde should report a value of $\epsilon$ for that wavelength. The fidelity of that model was tested with direct measurements of flux, discussed in the section to follow.

\section{Observations and Analysis}\label{observations}
 
The Polar Atmospheric Emitted Radiance Interferometer (P-AERI) is a Fourier-transform moderate-resolution ($1~{\rm cm}^{-1}$) infrared spectrometer which is sensitive in the $3.3~\mu{\rm m}$ to $20~\mu{\rm m}$ range. A two-channel detector utilizing photoconductive mercury cadmium telluride (HgCdTe) coupled with photovoltaic indium antimonide (InSb) in a sandwich configuration is employed, housed in a dewar cooled to 70 K using a Stirling-cycle cooler. This spectrometer is illuminated through an infrared-transmitting window via a reflective optic, in a housing outside in ambient air.  By means of that tipping mirror, sky spectra at zenith and calibration observations on either of two different blackbody sources (ambient and warm) are obtained. This instrument, and its later version with an extended-range of sensitivity (E-AERI) are described in more detail in \cite{Mariani2012}.

The P-AERI was operated by the University of Idaho from the Zero-altitude Polar PEARL Auxiliary Laboratory (0PAL), at 10 m elevation a.s.l., nearby the Eureka weather station. A near-continuous record with only a few short interruptions is available from fall 2006 through spring 2009, during which it was employed to measure the absolute downwelling infrared spectral radiance for atmospheric physics studies. This and the E-AERI instruments were co-located during one year either side-by-side at this site, or in concert from the PEARL facility roughly 15 km to the west of Eureka. The side-by-side data of several days were sufficient to allow verification of instrument specifications. The other observations were mostly in cloudy conditions but sensitive enough to show the presence of very thin ice-crystal clouds, which affect surface radiative forcing \citep{Mariani2012}, although those data were not used directly here. Nonetheless they are relevant to this analysis, as the careful tests show excellent agreement between instrument calibrations, within $1\%$ flux accuracy as expected in \cite{Knuteson2004}. A small positive radiance bias of P-AERI has been noted, and is being actively investigated, however, this would tend only to make those data slight overestimates of sky brightness (Walden, V.P. \& Turner, D.D. {\it private communication}). This may also be relevant for possible direct intercomparison of future sky brightness measurements. For example, P-AERI has since been redeployed to Summit Greenland.

\begin{figure}
\plotone{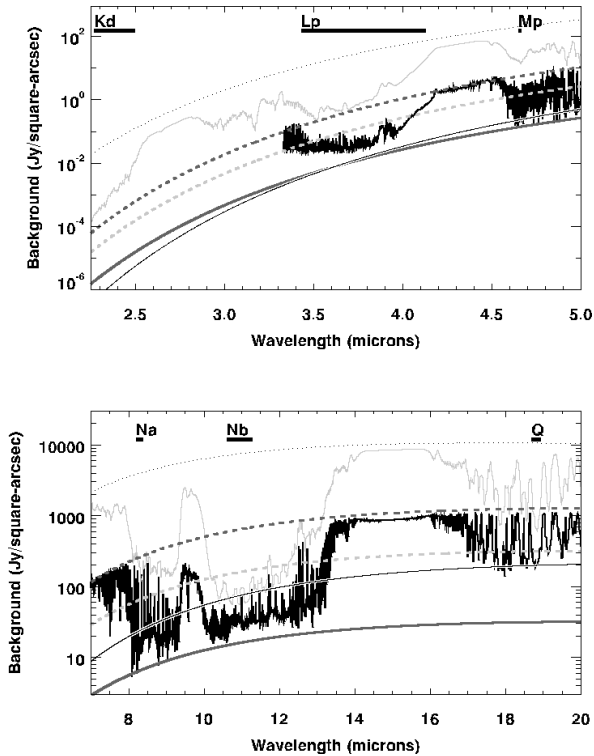}
\vspace{1 mm}
\caption{Median P-AERI spectrum. A grey curve indicates best theoretical model for the comparison temperate site; thick dark-grey dashed curve is for ``worst" Eureka conditions, light-grey for median, and solid grey curve for ``ideal." The thin black curve would be indicative of the typical conditions associated with South Pole; the same thermal model reported in \cite{Hidas2000}.}
\label{figure_spectra}
\end{figure}

Those periods of clear skies in darkness are of most interest here, in particular at times when simultaneous atmospheric temperature profiles are available. To exclude any complication from daytime conditions, only those periods between 21 October and 20 February of each winter were considered, when P-AERI operated from 0PAL. Processed spectra for those times were obtained from the the National Oceanic and Atmospheric Administration (NOAA) public archive\footnote{ftp://ftp1.esrl.noaa.gov/psd3/arctic/eureka}. This includes 449427 individual spectra, or samples of roughly once per minute. These were obtained from the archive after September 2015, when a fault in the high-frequency channel calibration was corrected, and all data reprocessed. 

The median of all those spectra is shown in Figure~\ref{figure_spectra}. Model limits are also plotted: the light grey shows the result for convolving transparency at 1.6 mm of PWV with thermal emission at $0~{\rm C}$ (Cerro Pachon, not shown, would be slightly higher) the thin black curve is equation 1 for $\epsilon=0.05$ and $T=-60~{\rm C}$.  Interestingly, it can be preferential at longer wavelengths to have a lower emissivity, even if temperatures are higher. The ``ideal" case for Eureka of $\epsilon=0.005$ and $T_{\rm peak}=-30~{\rm C}$ is shown as a thick dark-grey curve; a dashed dark-grey curve shows the ``worst" case of $\epsilon=0.20$. The light grey dashed curve is for values $\epsilon=0.05$ and $T_{\rm peak}=-22~{\rm C}$.

\subsection{Combined Dataset with Photometry}

Basic meteorological observations are obtained hourly at Eureka, including pressure, surface air temperature, cloud cover, wind, and relative humidity. The last is essentially always near saturation in winter.  Calm winds have been shown to correlate with good sky conditions at PEARL, but a more useful criterion here is the visual inspection of sky clarity from sea level. Although the observer's assessment of sky-clarity does not conform to a standard familiar in astronomy, usually expressed in eigths, best conditions do essentially correspond to what would be familar as the best 1/8th elsewhere. A determination of ``clear" by the meteorological observer denotes the complete absence of visible ice crystals in the atmosphere, which is known to correspond to truly photometric conditions at PEARL \citep{Steinbring2012}. This condition was reported in the hour immediately preceding 46148 spectral records, or about 10\% of the time.  That is not to say that other times were not clear, but this visual confirmation verifies those cases when only atmospheric thermal emission was likely to be important.

\begin{figure*}
\plottwo{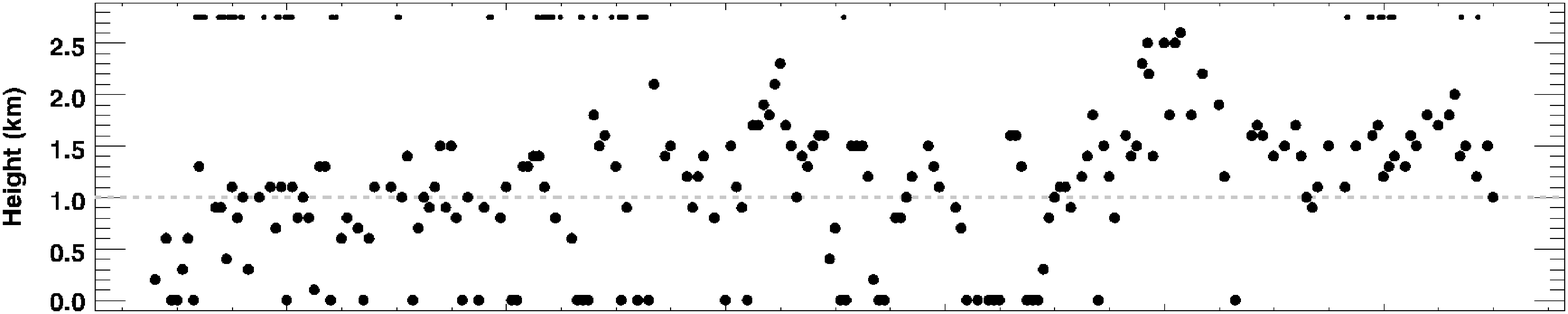}\\
\plottwo{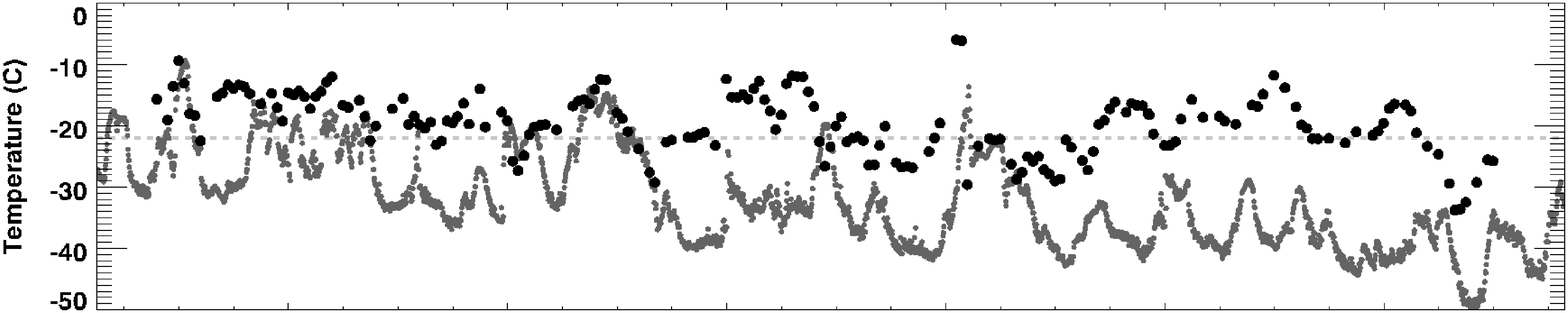}\\
\plottwo{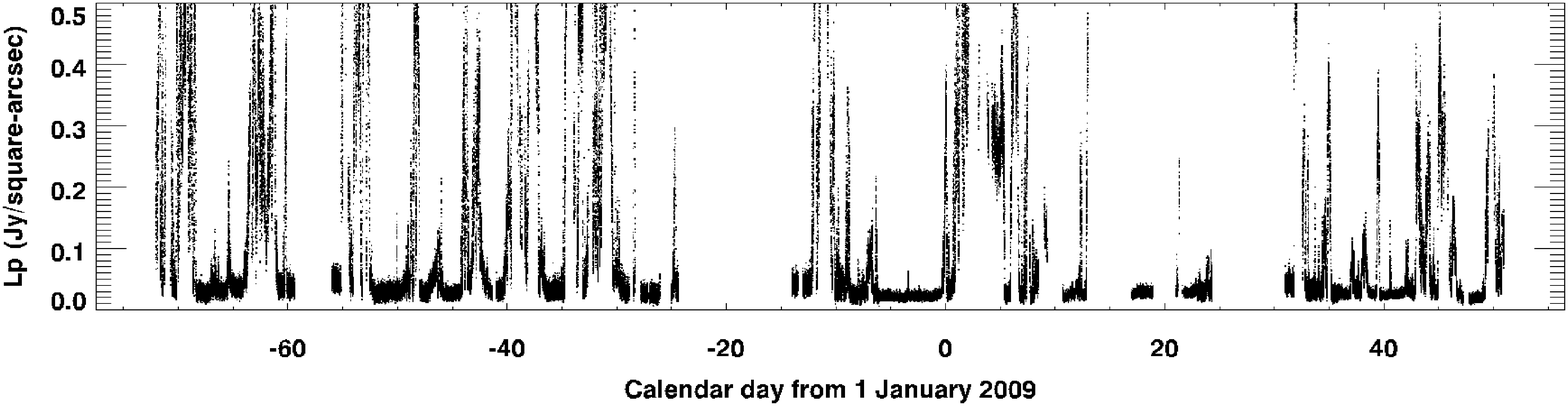}
\vspace{1 mm}
\caption{Sky-brightness data in the most transparent portion of $L_{\rm p}$ band for one winter dark period given in calendar days starting from 1 January 2009; other bands are similar, as are the further two winters studied. The two upper panels show contemporaneous inversion and weather conditions: peak height (top panel) and temperature; ground temperature (grey points), and mean values as horizontal dashed grey lines. Clear periods visually verified from sea level are indicated by dark filled circle at the very top of the panels.}
\label{figure_data}
\end{figure*}

For each P-AERI spectrum, photometry was performed in the most transparent portions of $L_{\rm p}$, $M_{\rm p}$, $N_{\rm a}$, $N_{\rm b}$, and $Q$ band averaged over a $0.2~\mu{\rm m}$ bandpass centred at the wavelength specified in Table 1. The season in winter 2008/09 illustrates the available data, and is shown in Figure~\ref{figure_data}. The other two seasons are similar.  A filled circle at the upper-most panel indicates those times when skies were visually confirmed to be clear. The top panel plots the height of the inversion peak, as measured from the radiosonde data. A spline was fit to the data to retrieve this; these datapoints are those for which there is at least one associated P-AERI spectrum, that is, taken within 30 minutes of the launch. This includes some 15645 spectra, or about 3.5\% of the total sample.  The panel second-to-top gives the surface temperature from Eureka weather station records (dark grey points), and the peak temperature obtained from the radiosonde launches (dark circles).  Light-grey horizontal dashed lines indicate the mean value of ${\bar T}_{\rm peak}=-22~{\rm C}$ and ${H}_{\rm peak}=1.0~{\rm km}$.

\subsection{Identifying Clear-Sky Conditions}

\begin{figure}
\plotonenarrow{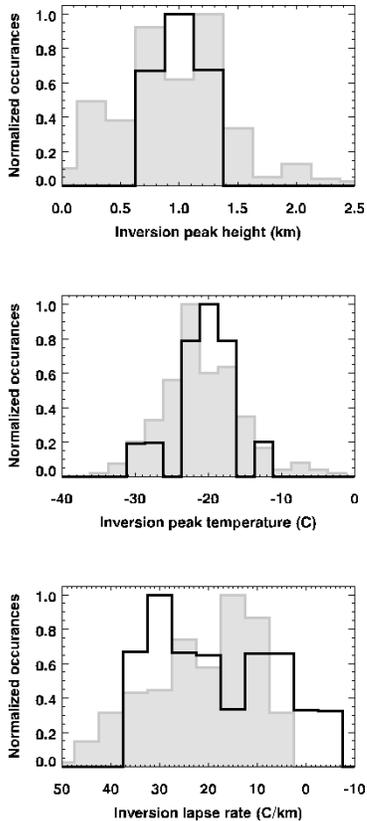}
\vspace{1 mm}
\caption{Normalized distributions of peak height, temperature, and associated inversion lapse rate for Eureka during winter darkness but not confirmed clear (grey shading) and exclusively times when skies were confirmed clear (dark outline). Note how conditions differ under clear skies, particularly the shift to high lapse rate and a peak height more tightly centered near 1 km.}
\label{figure_peak_distributions}
\end{figure}

Photometric conditions in the optical from PEARL correspond directly with the lack of ice crystals in the lower atmosphere, which are known to occur about 50\% of the time \citep{Steinbring2012}. So visual confirmation of clear skies from sea level coincident with a radiosonde provides a check on the correctness of the simple model prescription of thermal sky brightness for PEARL.  Figure~\ref{figure_peak_distributions} gives the distributions of associated inversion peak height, temperature, and lapse rate for every profile obtained. Dark outlines are for confirmed clear-sky conditions only; grey shading are other times. Importantly, under clear skies there is a well-defined range of inversion peak height, confined within 0.5 km to 1.5 km. See also the narrow distributions of peak temperatures at all times, and how the lapse-rate distribution is notably different for clear skies than otherwise. This confirms that clear skies are associated with a strong inversion, and that the breakdown of the inversion is more likely to be associated with cloud.

\subsection{Extraction of Emissivity}

\begin{figure*}
\plottwo{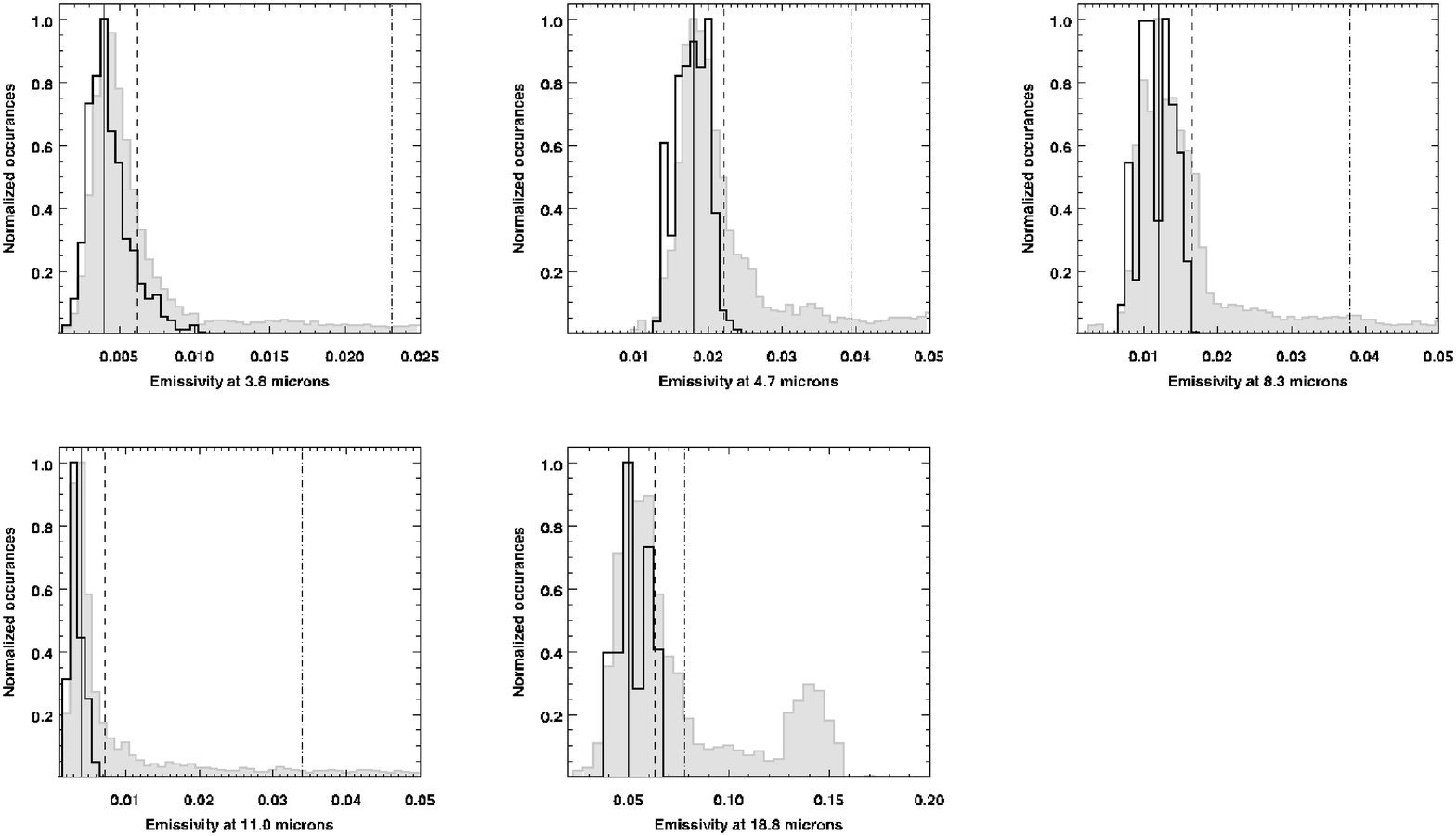}
\vspace{1 mm}
\caption{Histograms of inferred emissivity at given wavelengths in the 5 bands under study (grey shading). Results for those times which were known to be clear simultaneously with a radiosonde profile are indicated by the dark outline. Measured mode, median, and means are indicated by vertical solid, dashed, and dot-dashed lines, respectively.}
\label{figure_emissivity_distributions}
\end{figure*}

The coincidence of radiosonde profiles and P-AERI brightness measurements from sea level allows a straightforward measure of emissivity across all of the bands in question.  This follows from inverting equation 1, and directly reporting emissivity for the observed $T_{\rm peak}$; results are plotted in Figure~\ref{figure_emissivity_distributions}. The light-grey shaded regions are all data, and the dark outlines indicate those observations obtained when it was known to be clear in Eureka.  Note how similar these results are to the expectations from the atmospheric transmission model of Section~\ref{model}, particularly under clear skies; a range of emissivities consistent with $\epsilon=0.005$ to $0.05$, typically. The cause of the secondary ``bump" towards higher emissivity at $18.8~\mu{\rm m}$ is unknown, but was evidently not associated with photometric conditions. A hard upper limit of $\epsilon=0.20$ is also consistent with that used by \cite{Hidas2000} in characterizing South Pole infrared backgrounds.

Although instrumental error on individual fluxes per band is near 1\%, uncertainties in recovering the instantaneous peak inversion temperature, and thereby an estimate of emissivity, are necessarily larger.  This is less easy to characterize per sample because it must involve error in both the temperature and pressure measurements ($0.5~{\rm C}$ and $25~{\rm Pa}$ at each elevation) along with the fidelity of their profile fits. That is not precisely the same as the true effective sky temperature, and does not account for temporal fluctuation during a balloon flight.  Systematic error in predicting the sky brightness at PEARL can still be constrained though. Over the ensemble, this is conservatively well within the $4~{\rm C}$ peak inversion temperature standard deviation, which scales flux to within 20\%, declining to 5\% as the observing wavelength tends to $20~\mu{\rm m}$.  Knowledge of true effective sky temperature is plausibly within $2~{\rm C}$, similar to the difference in surface temperature at 600 m and warmest inversion peak; a global 10\% uncertainty in emissivity is a reasonable guide, as well as to characterize the real sky brightness distribution, which follows.

\subsection{Brightness Distributions and Comparison}

\begin{figure*}
\plottwomedium{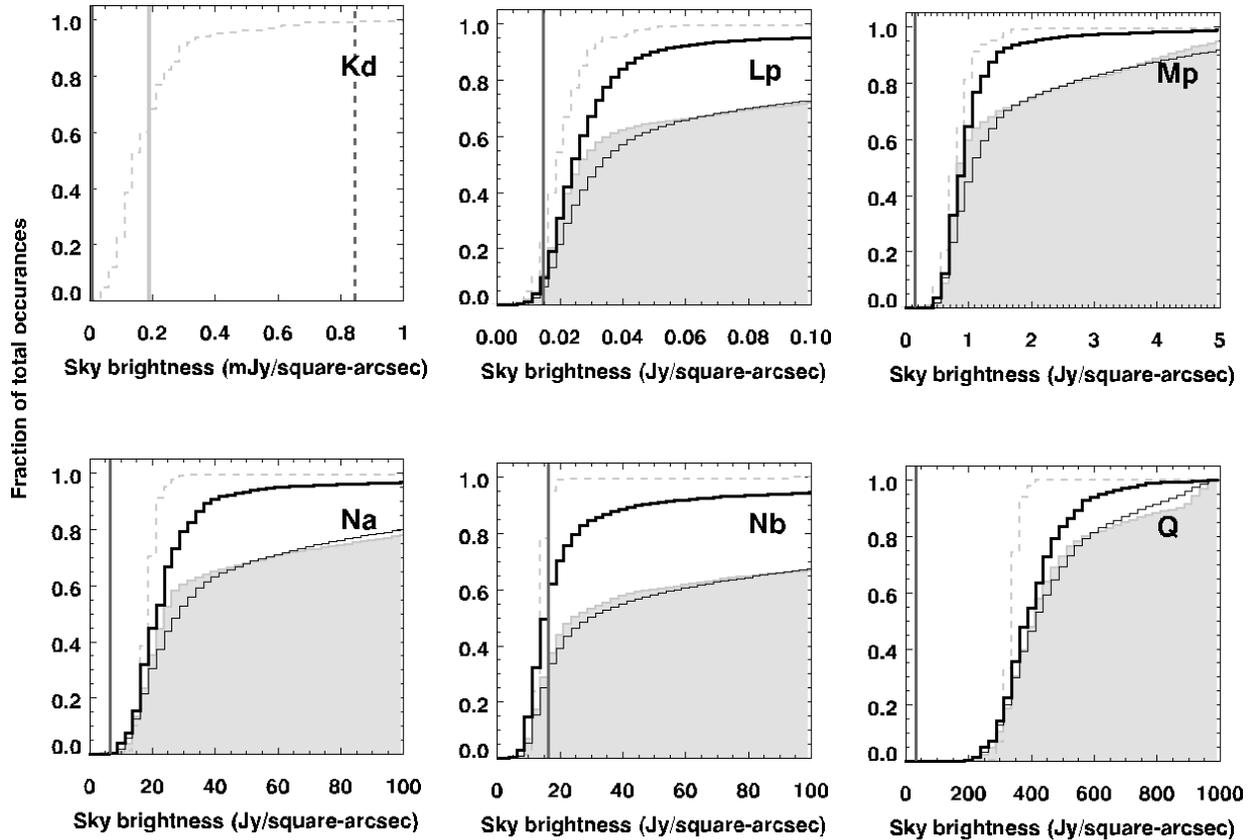}
\vspace{1 mm}
\caption{Normalized cumulative sky-brightness distributions observed with P-AERI (all data: thin dark outline; confirmed clear: thick dark outline). Results of models are also shown: grey shading outlines a distribution of inferred emissivity, and dashed outline instead the observed range of peak temperature. Vertical dark-grey dashed line indicates ``worst" case for $K_{\rm dark}$, solid vertical lines are ``ideal."}
\label{figure_brightness_distributions}
\end{figure*}

Distributions of observed brightness in bands $L_{\rm p}$, $M_{\rm p}$, $N_{\rm a}$, $N_{\rm b}$, and $Q$ were calculated. Results are plotted in Figure~\ref{figure_brightness_distributions}.  For comparison, the grey curves outline the simple model of Section~\ref{model} in two variants. These simulated values are averaged over a $0.2~\mu{\rm m}$ bandpass, centred at the observed wavelength. By first taking the measured modal emissivity in each band for $\epsilon$, and using the distribution of peak temperatures $T_{\rm peak}$ in equation 1 the result is shown as a light-grey dashed curve.  Note how closely this matches the median in observed sky brightness, although it appears somewhat too steep in this cumulative histogram. A second, alternate variant of this simple model is to instead use the mean peak temperature over all radiosonde launches, and apply the measured distribution of emissivities in each band. This result is shown as a light-grey shaded region, which together with the first model bracket the clear-sky conditions. That the second of these two models is a good fit to the brighter parts of the distributions confirms that high values of emissivity (those rarer times associated with thick cloud) are those times associated with higher sky brightness.

The simple model also helps compare thermal background between different sites under photometric skies: at one limit is an idealized sky brightness, taken to be equation 1 with the ``ideal" emissivity of $\epsilon=0.005$ and $T_{\rm peak}=-30~{\rm C}$; shown as a thick, vertical grey line in Figure~\ref{figure_brightness_distributions}. Note how this emphasizes the sharp, dark edge of the $L_{\rm p}$ sky emission. That is expected, as it is near the wavelength at which this ``ideal" case overlaps that of mean transparency, but coldest conditions (see Figure~\ref{figure_spectra}).  In other words, this is where the advantage of colder conditions can begin to overcome poorer transparency, which also explains how the extreme temperatures of South Pole (down to $-60~{\rm C}$) can result in lower emission, even when $\epsilon=0.05$. The median and mean sky brightnesses in each observed band are tallied in Table~\ref{table_results}. Not surprisingly, there is less difference between Arctic and Antarctic at long wavelengths, but improvement over a high mid-latitude site at short wavelengths becomes significant. For example, using $\epsilon=0.05$, the mean model value for Cerro Pachon at $L_{\rm p}$ is $2~{\rm Jy}~{\rm arcsec}^{-2}$; Maunakea is $1~{\rm Jy}~{\rm arcsec}^{-2}$, rising to $6~{\rm Jy}~{\rm arcsec}^{-2}$ at $M_{\rm p}$. In the $N$ window, near $8.9~\mu{\rm m}$ \cite{Chamberlain2000} report South Pole has a nighttime median of $50~{\rm Jy}~{\rm arcsec}^{-2}$, about the same as found at Eureka with P-AERI. Although the data are for austral summer temperatures, estimates at Dome C do not improve greatly on this at longer wavelengths: there $N_{\rm b}$ reaches $43~{\rm Jy}~{\rm arcsec}^{-2}$ and $310~{\rm Jy}~{\rm arcsec}^{-2}$ at $Q$ \citep{Walden2005}, which are not quite a factor of two better.

Finally, a confident extrapolation shortward into $K$ band can be made. A conservative upper bound on sky brightness is obtained from assuming a ``worst" case of $\epsilon=0.20$ and $T_{\rm peak}=-18~{\rm C}$; shown as a thick dashed grey vertical line in Figure~\ref{figure_brightness_distributions} in the panel depicting $K_{\rm d}$.  Together with the measured distribution of inversion temperatures, this bounds the full range of thermal sky brightness in the $K_{\rm d}$ band; clear skies are almost never brighter than $850~\mu{\rm Jy}~{\rm arcsec}^{-2}$, with a median of $170~\mu{\rm Jy}~{\rm arcsec}^{-2}$ (vertical light-grey line) for $\epsilon=0.05$. This is a dramatic improvement over temperate sites. For comparison, at $2.4~\mu{\rm m}$ thermal sky background is then about $4~{\rm mJy}~{\rm arcsec}^{-2}$ at Gemini North, and $8~{\rm mJy}~{\rm arcsec}^{-2}$ for Gemini South. Here PEARL would be a order of magnitude darker, on average, and possibly as much as a factor of 40.  In fact, this is comparable to the measured median and third-quartile values at South Pole between $155$ and $270~\mu{\rm Jy}~{\rm arcsec}^{-2}$, although not quite as low as the best quartile of $80~\mu{\rm Jy}~{\rm arcsec}^{-2}$ quoted by \cite{Lawrence2002}.

\begin{deluxetable}{lrccc}
\tablecaption{Median and Mean Sky Backgrounds\tablenotemark{a} with Associated Atmospheric Emissivities\label{table_results}}
\tablewidth{0pt}
\tabletypesize{\small}
\tablehead{\colhead{Bandpass} &\colhead{$\lambda$ ($\mu{\rm m}$)} &\colhead{$F_{\rm median}$} &\colhead{$\bar{F}$} &\colhead{$\bar{\epsilon}$}}
\startdata
Kd\tablenotemark{b}  &2.4  &$1.7\times10^{-4}$ &$3.1\times10^{-4}$ &0.050\\
Lp                   &3.8  &$3.6\times10^{-2}$ &$1.3\times10^{-1}$ &0.023\\
Mp                   &4.7  &$1.2$              &$2.0$              &0.039\\
Na                   &8.3  &$31$               &$64$               &0.038\\
Nb                   &11.0 &$32$               &$130$              &0.034\\
Q                    &18.8 &$460$              &$550$              &0.078
\enddata
\tablenotetext{a}{In units of ${\rm Jy}~{\rm arcsec}^{-2}$.}
\tablenotetext{b}{Results based on the model in this paper.}
\end{deluxetable}

\section{Summary and Conclusions}\label{summary}

Archival measurements of the thermal infrared downwelling flux above Eureka, on Ellesmere Island Canada, have been presented. Three complete winter seasons sampled typical variation in thermal inversion temperature. Available meteorological data have been compared with other atmospheric studies which verify that reliably cold, clear conditions for the PEARL site at 600 m elevation will result in dark skies in the infrared during winter - the first time this has been shown for a High Arctic mountain site. Contemporaneous visual estimates of sky clarity with twice-daily balloon-borne radiosonde profiles of air temperature allow extraction of the atmospheric transmission in $L_{\rm p}$, $M_{\rm p}$, $N_{\rm a}$, $N_{\rm b}$, and $Q$ bands. 

A simple model of thermal emission provides a good general fit to these data, and allows extrapolation into $K$ band. Skies at $2.4~\mu{\rm m}$ should be particularly dark, more than an order of magnitude better than for current mid-latitude infrared observatories; Gemini North and South are used for comparison here as they analogously allow access to both hemispheres. Darkness is comparable to South Pole under clear skies. Unfortunately, $2.4~\mu{\rm m}$ is shortward of P-AERI sensitivity, which sampled at zenith only.  Direct confirmation in $K_{\rm d}$ with a dedicated instrument would be desirable, especially characterization of airmass dependence. This will be influenced by the aerosol content of the lower atmosphere, in particular the vertical distribution of diamond dust, which differs from the Antarctic plateau.

Beyond the important primary result of dark skies in the thermal infrared, knowledge of the other observing conditions for PEARL permits a secondary inference to be drawn. The rooftop observing platform of PEARL can provide excellent seeing under calm, clear conditions, with a median under $0\farcs76$ in $V$ at 8 m elevation \citep{Steinbring2013}. So relative to South Pole, this constitutes a significant improvement, as it is well known that optical seeing within its strong inversion layer is poor: $1\farcs9\pm0\farcs6$ average and standard deviation in $V$ at 7.5 m elevation \citep{Travouillon2003}. For a given point source with the same sky brightness this is typically a gain of $S/N\approx2.5$ for PEARL due to allowing a smaller, optimized photometric aperture, or effectively a gain of a magnitude in depth for the same exposure time. In AB magnitudes $200~\mu{\rm Jy}~{\rm arcsec}^{-2}$ corresponds to $18.1~{\rm mag}~{\rm arcsec}^{-2}$, less than a magnitude brighter than best conditions at South Pole. So the implication is that the disadvantage of warmer conditions at PEARL relative to South Pole is counterbalanced by better photometry. The combination of the two sites could be very powerful in the era of long-term time-domain surveys, as it exploits their offset observing seasons and particularly dark skies in $K$ - together providing greatly improved efficiency over temperate sites. One program is already anticipating a survey employing twin $K_{\rm d}$-optimized telescopes for South Pole and PEARL to take advantage of this ``bi-polar" astronomical opportunity \citep{Moore2016}.

\acknowledgements

I would like to acknowledge Von Walden for both a careful review of this manuscript and helpful advice regarding AERI data, together with Penny Rowe and Chris Cox for their support of the P-AERI database. The Canadian Network for the Detection of Atmospheric Change (CANDAC) is responsible for operational support of the PEARL and 0PAL facilities, including this instrumentation. I thank Anna Moore and Michael Ashley for helpful conversations about sky brightness, and Suresh Sivanandam for discussions on infrared instrumentation, in particular for pointing out available Eureka data. Thoughtful comments by an anonymous referee helped improve the original manuscript. Eureka aerosonde profiles used here are from the Integrated Global Radiosonde Archive of the National Oceanic and Atmospheric Administration (NOAA); P-AERI data were obtained using the NOAA public FTP data service.  Eureka surface meteorological data are courtesy of Environment Canada, obtained from Historical Climate Data online archive.

\end{document}